\pdfoutput=1
\documentclass[11pt]{article}

\usepackage[]{acl}

\usepackage{times}
\usepackage{latexsym}
\usepackage{booktabs}
\usepackage{enumitem}

\usepackage[T1]{fontenc}
\usepackage[utf8]{inputenc}
\usepackage{multirow}
\usepackage{multicol}
\usepackage{lipsum}
\usepackage{CJKutf8}
\usepackage{makecell}

\usepackage{amsmath}
\usepackage{microtype}

\usepackage{inconsolata}

\usepackage{graphicx}

\title{One-Step Retrieval Framework for Real-Time Sponsored Search Ads \\ Using Hierarchical Text Representations}

\author{
  \textbf{Tongtong Liu},
  \textbf{Renyu Zhang},
  \textbf{Jiayu Ding},
  \textbf{Hongchao Guo},
  \\
  \textbf{Xintao Yang},
  \textbf{He Wei},
  \textbf{Zhaoyu Li},
  \textbf{Haiyang Wu}
\\
 Tencent Inc.\\
   \{uniqueliu,yurenzhang,garryding\}@tencent.com \\ 
   \{peterrguo,xintaoyang,whywei,joeyzyli,gavinwu\}@tencent.com
}

\begin{document}
\maketitle
\begin{abstract}

Traditional retrieval systems typically use multi-stage cascading architectures (MCA), where each module is optimized independently, leading to inconsistent objectives and the premature elimination of high-potential candidates. 
Recent LLM-based generation methods offer end-to-end solutions but use discrete semantic identifiers (SIDs) to retrieve ads, which are not learned by the base LLM and require memorization of numerous SID-to-ad mappings during SFT, suffering from limited generalization to unseen ads, high maintenance and update costs. The one-to-one mapping between SIDs and advertisements leads to inefficient decoding. Moreover, these methods rely on a small reward model (e.g. pctr) for relevance and ranking, limiting the LLM’s ability to fully assess ads’ commercial value. 
To address these challenges, we propose \textbf{A} u\textbf{N}ified \textbf{G}eneration-discriminative-ranking rea\textbf{L} time r\textbf{E}trieval (\textbf{ANGLE}) framework. ANGLE uses LLM-generated hierarchical  textual representations, which consist of commercial intent that provide high-level overviews and ad abstract that deliver fine-grained details. Additionally, ANGLE integrates retrieval, relevance, and ranking directly within a single LLM, enabling precise and efficient ranking of ads by leveraging the full capabilities of the LLM. We applied ANGLE to the real-world search scenarios, achieving a 1.81\% increase in consumption and a 2.16\% increase in gross merchandise volume (GMV).  We also conducted offline evaluations of ANGLE and seven baselines, with ANGLE outperforming all across key metrics such as HR and ACR.

\end{abstract}

\section{Introduction}

% 传统架构的问题
Ad retrieval systems efficiently retrieve and display high-value ads that match users' search intent from large-scale corpora. Retrieval, relevance, pre-ranking and ranking constitute four essential components of such a system, addressing user search intents, advertiser delivery requirements, and platform revenue objectives. Traditional retrieval systems typically use multi-stage cascading architectures (MCA), in which candidates are sequentially passed through retrieval, relevance, pre-ranking, and ranking component, with each stage taking the previous stage’s output as input, and finally selecting the top-k ads for display. However, this approach faces several drawbacks. Firstly, errors or omissions in earlier stages are propagated to subsequent ones, limiting the overall quality of the final results. Secondly, the modules are often trained independently, lacking joint optimization, which makes it difficult to achieve a globally optimal solution.  Thirdly, processing a large number of candidates at each stage leads to high computational costs and slower response times. In recent years, several studies \cite{fei2021gemnn, huang2023cooperative, zhang2023rethinking} have been conducted to address the aforementioned challenges of MCA. However, these methods still fail to break through the framework of  MCA.

With the rapid development of LLM, generative approaches have recently been explored for directly producing candidates for users. Most existing methods\cite{chen2025onesearch, deng2025onerec} employ discrete semantic identifiers (SIDs), typically generated via RQ-VAE\cite{lee2022autoregressive} to represent candidate docs/ads. Upon receiving a user query, the LLM generates a set of SIDs conditioned on the query, which are subsequently used to retrieve relevant ads that are presented to the user. To further refine the recommendation process, lightweight reward models—such as click-through rate (CTR) predictors—provide feedback on the relevance and effectiveness of the displayed ads.

However, existing work suffers from two main \textbf{limitations}. Firstly, ads are represented by discrete SIDs \cite{tay2022transformer, fu2025forge} that rely on unique discrete IDs to characterize each ad. Consequently, the LLM is required to learn the mapping between SIDs and ads from scratch, which limits its ability to fully leverage its inherent text generation capabilities and reduces the overall interpretability of the model. Furthermore, these mappings exhibit poor generalization, necessitating the creation of new mappings for previously unseen docs/ads. Secondly, current methods depend on one or more lightweight models(e.g., predicted click-through rate predictors) as reward models to guide the optimization of the LLM, which constrains the full potential and capabilities of the LLM. Therefore, developing more effective advertising representations and reward mechanisms is essential for fully leveraging the capabilities of LLMs within the one-step generation framework.

To address these challenges, we propose \textbf{A} u\textbf{N}ified \textbf{G}eneration-discriminative-ranking rea\textbf{L}-time r\textbf{E}trieval framework named \textbf{ANGLE}, employs a hierarchical semantic text \textemdash intent and abstract pair \textemdash to represent advertisements (ads), integrating relevance and ranking objectives directly into the LLM training process to achieve end-to-end retrieval. We incorporate both discriminative and ranking capabilities into the base LLM, resulting in the generation-discriminative-ranking LLM (GDR-LLM). Specifically, ANGLE leverages base LLM to generate intents (broad, intent-expressive text) and abstracts (precise, ad-derived text) for ads, subsequently constructing an inverted index mapping intent-abstract (int-abs) pairs to the associated ads. Upon receiving a user query, the GDR-LLM first infers the query intent, then generates the corresponding ad abstract aligned with this intent, and retrieves the relevant ads via the aforementioned inverted index. These retrieved ads are subsequently sent directly to the ranking stage for processing, after which they are displayed to users.

% introduction还有空间，可以突出一下ANGLE和其他工作相比的优点
The primary \textbf{innovation} of this work lies in utilizing hierarchical semantic text as an intermediate representation and designing a GDR-LLM to enable end-to-end retrieval from users to ads.  Specifically, we represent user search intents and advertising needs as int-abs pairs, where both components are concise texts conveying rich semantic information. This semantic text representation allows ANGLE to fully utilize the world knowledge embedded in LLM, as it aligns well with their generative mechanisms.  Furthermore, our approach demonstrates strong generalization capabilities: for new ads, the int-abs pair can be efficiently updated by generating it directly from ad textual content. This representation also offers high interpretability, which facilitates clear understanding and analysis. In addition, GDR-LLM incorporates raw relevance evaluation and commercial ranking data into its supervised fine-tuning process, thereby generating highly commercial advertisements that align with relevance criteria. Unlike recommendation scenarios, search scenarios require much higher precision and relevance. With enhanced relevance and ranking capabilities, GDR-LLM enables the simplification of the MCA pipeline: retrieved ads are sent directly to the re-ranking stage and subsequently displayed to users, achieving end-to-end retrieval.

% The main contributions of our work are as follows: (1) We propose ANGLE, an end-to-end, real-time, extreme-scale framework, and demonstrate its application in a real-word system operating at the tens-of-millions scale. (2) We propose representing advertisements using hierarchical semantic text, addressing the issues of lack of interpretability and poor generalization associated with the use of SIDs in prior architectures. (3) We propose a Generate-Discriminate-Rank LLM that integrates both discriminative and ranking functionalities within a unified framework, fully leveraging the capabilities of LLM. (4) We validated the effectiveness of ANGLE in real-world scenarios through A/B testing, achieving a consumption increase of 1.81\%, a GMV growth of 2.16\%, and a 1.5\% rise in the number of clicks.

% whywei 0508: 格式化贡献，审稿人一眼看到主要工作
The main contributions of our work are summarized as follows:
\begin{enumerate}[
    leftmargin=1em, labelsep=0.5em, topsep=0ex,
    itemindent=0em, listparindent=0em, parsep=0ex, itemsep=0ex,
    ]
    \item[-] \textbf{Framework}: We propose ANGLE, an end-to-end real-time framework validated at tens-of-millions scale in production.
    \item[-] \textbf{Representation}: We introduce hierarchical semantic text for ad representation, addressing interpretability and generalization issues of SIDs.
    \item[-] \textbf{Model}: We design a generation-discriminative-ranking LLM that integrates discrimination and ranking capabilities within a unified framework, fully leveraging the capabilities of LLM.
    %\item[-] \textbf{Evaluation}: We demonstrate significant improvements in real-world metrics: 1.81\% increase in consumption, 2.16\% increase in GMV, and 1.5\% increase in clicks. by ltt 0508,评估添加离线
    \item[-] \textbf{Evaluation}:  We conducted both online A/B testing and offline experiments to evaluate the effectiveness of the ANGLE. The A/B testing yielded a 1.81\% increase in consumption, a 2.16\% increase in GMV, and a 1.5\% increase in clicks. Additionally, offline comparisons against seven baselines show that ANGLE achieved the best performance across HR, MAP, and ACR metrics.
  
\end{enumerate}

\begin{figure*}[t]
  \centerline{\includegraphics[width=1.0\linewidth]{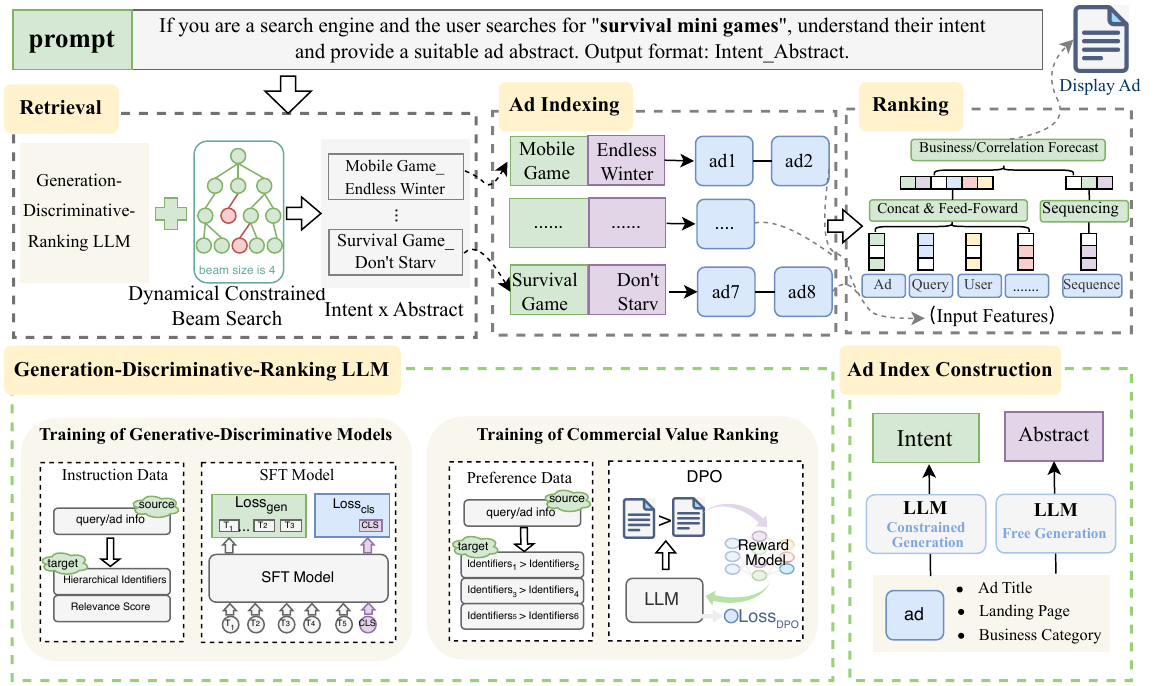}}
  \caption {The overall architecture of ANGLE framework, comprising retrieval, ad indexing and ranking. After the user inputs a query, the retrieval employs a GDR-LLM to produce valid intent-abstract (int-abs) pairs through dynamic constrained decoding. The Ad Indexing module then retrieves relevant ads based on the generated int-abs pairs.The retrieved ads are directly sent to the ranking stage for sorting before being presented to the user.
  \label{fig:LOGR}
}
\end{figure*}

\section{Related Work}

% whywei 0508: 概括相关工作的具体内容，方便审稿人挑自己不熟悉的领域查看
We review related work from three perspectives: traditional multi-stage cascading architectures, generative retrieval approaches, and document representation methods in LLM-based retrieval systems.

\subsection{Multi-stage Cascading Architectures}
Multi-stage cascading architectures (MCA) \cite{covington2016deep, wang2011cascade} are widely used in industrial advertising systems to balance computational efficiency and ranking accuracy. This approach divides retrieval pipeline into sequential stages\textemdash typically retrieval\cite{freymuth2025hierarchical}, relevance\cite{dey2025middleman}, pre-ranking\cite{zhao2025hybrid}, and ranking\cite{huang2025towards} \textemdash where the output candidates from each stage serve as the input for the next.%The retrieval module filters ads from hundreds of millions of candidate ads to find those that match the user's search intent, while relevance filters out irrelevant ads. The pre-ranking module then sorts these ads based on relevance and commercial value, and after that, the ranking module performs detailed bid-based sorting and presentation.
However, our proposed ANGLE framework eliminates relevance and pre-ranking, allowing generated content to proceed directly to ranking and thus significantly simplifying the retrieval pipeline.

\subsection{Generative Retrieval}

Generative retrieval \cite{zhou2025openonerec} reformulates the retrieval problem as a generation task, whereby the model directly generates the content \cite{li2023multiview} or SIDs \cite{zhang2026onemall} of relevant docs/ads to accomplish retrieval. %Specifically, given a natural language query, the generative model directly produces a set of representations that are used to retrieve relevant doc/ads, enabling the end-to-end retrieval process.
For example, UniSearch \cite{chen2025unisearch} generates SIDs for video, retrieving relevant videos by producing SIDs based on user queries. RARE \cite{liu2025real} utilizes commercial intent (CIs) as the representation of ads to enable one-step query-to-ad retrieval.

\subsection{Doc Representation}

LLM transforms the retrieval into a generation task by generating semantic representations of doc/ads. Existing semantic representations are mainly divided into three categories:  semantic identifiers (SIDs), semantic term sets (STerms), and semantic compressed text (SText). SIDs are discrete semantic representations generated by encoding advertising content using models such as RQ-VAE\cite{lee2022autoregressive}. For instance, these works \cite{liu2025onerec, zhang2025gpr} employs SIDs to effectively represent ads. STerms refer to a set of representative terms extracted directly from the content of candidates. For example, SEAL \cite{bevilacqua2022autoregressive}represents documents using n-grams extracted from the text. SText uses semantically compressed text to represent doc/ads. For example, RARE\cite{liu2025real} uses the commercial intent to represent the advertisement, and GRAM \cite{pang2025generative} performs product retrieval from queries using code-based implementation. Unlike the aforementioned work, we propose a novel hierarchical text representation approach to enable generative retrieval for ads.

\section{Method}

In this section, we introduce in detail the proposed ANGLE framework, and the two components of ANGLE: ad indexing and retrieval.

% 这里的量级需要再确认
\begin{figure}[t]
  \centerline{\includegraphics[width=1.0\linewidth]{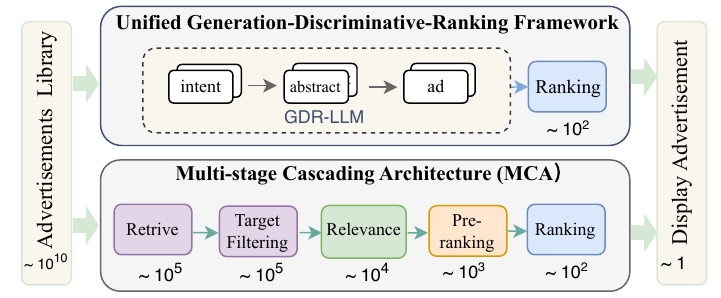}}
  \caption {Comparison of ANGLE and MCA: ANGLE uses a LLM to directly generate hundreds of candidate ads for ranking, streamlining the retrieval pipeline.
  \label{fig:figure_2}
}
\end{figure}

\subsection{End-to-end Generative Framework}

In this paper, we propose \textbf{A} u\textbf{N}ified \textbf{G}eneration-discriminative-ranking rea\textbf{L}-time r\textbf{E}trieval (\textbf{ANGLE}) framework that directly generates high-quality ads based on user queries and sends these candidate ads directly to the ranking. In the ANGLE framework, an integrated generation-discriminative-ranking LLM (GDR-LLM) is designed to produce ads that balance relevance and commercial value, enabling direct access to the final ranking stage as shown in the Figure \ref{fig:LOGR}. Additionally, the framework employs semantic representations at varying granularities, ranging from coarse-grained intent overviews to fine-grained abstracts, to enhance retrieval accuracy and indexing efficiency. More detailed information about int-abs pairs will be provided in Section ~\ref{int-abs}.

The retrieval process is shown in the Figure \ref{fig:LOGR}, when ANGLE receives a user query, the GDR-LLM produces int-abs pairs with high commercial value and relevance. This phase is called \textbf{retrieval}. After obtaining int-abs pairs, ANGLE queries the ad index to retrieve the corresponding ads, which corresponds to the \textbf{ad indexing} stage. Once the ads are acquired, ANGLE bypasses the traditional relevance and pre-ranking modules and directly sends the ads to the ranking stage. The design of the GDR-LLM, as well as the details of the retrieval, ad indexing, and ranking components, will be thoroughly discussed in the following sections.

%The end-to-end generation framework streamlines the advertising pipeline by directly producing high-quality, commercially valuable candidate ads from user search queries and immediately advancing them to the fine-grained ranking stage. By leveraging a LLM, it effectively mitigates error propagation and resolves the optimization objective inconsistencies inherent in traditional multi-component architectures (MCA), resulting in a more efficient and coherent ad generation process. 

\subsection{Generation-Discriminative-Ranking LLM}
This section focuses on the training process of the generation-discriminative-ranking LLM (GDR-LLM), which combines generation, discrimination, and ranking. As shown in the Figure \ref{fig:LOGR}, the GDR-LLM relies primarily on three loss functions for training: $L_{gen}$, $L_{dis}$, and $L_{dpo}$. The following sections describe the model design and data composition of these three components.

\paragraph{Learning to Generate. }
The objective of generation is to produce natural language representations of user search intents and advertising abstracts. The training data primarily originate from online user click data consisting of query-ad pairs. The organization of SFT data will be described in detail in the “Training Data” in section ~\ref{traindata} and ~\ref{setting}.
%We input advertising content into LLM (e.g., ChatGPT) to generate high-quality  commercial intent  and corresponding advertisement abstracts. Through this process, we construct a dataset of approximately 2 million query-intent-abstract pairs.
%These data are subsequently used to fine-tune our base model.
During training, we employ the cross-entropy loss function to measure the discrepancy between the predicted probability distribution and the ground-truth labels, as formally defined  as follows,

\begin{equation}
%\begin{split}
L_{gen} = - \sum\limits_{t=1}^T logP(y_{t} | y_{<t}, x)
%\end{split}
\end{equation}

where T is the number of generated tokens, $y_{t}$ is the t-th token in the target sequence, $y_{<t}$ represents all tokens before the t-th token, x is the model input and $P(y_{t} | y_{<t}, x)$ is the probability that the model correctly predicts token $y_{t}$ given the previously generated tokens and the input x.

\begin{table*}[]
\begin{tabular}{c|cc}
\toprule
\textbf{Ad Information}  & \textbf{Commercial Intents}       & \textbf{Ad Abstract}      \\  \midrule
\multicolumn{1}{c|}{\multirow{3}{*}{\begin{tabular}[c]{@{}l@{}}\textbf{Title}: In Endless Winter, as long as the fire \\
 burns well, there are no worries about life  \\
in the apocalypse!\end{tabular}}} & \multicolumn{1}{c}{Role-playing Games}            & \multicolumn{1}{c}{Endless Winter} \\ %\cline{2-3} 
\multicolumn{1}{c}{}                                                                                                                                                                                                                      & \multicolumn{1}{|c}{Post-apocalyptic Mobile Games} & \multicolumn{1}{c}{Endless Winter} \\ %\cline{2-3} 
\multicolumn{1}{c}{}     & \multicolumn{1}{|c}{Survival Challenge Game}       & \multicolumn{1}{c}{Endless Winter} \\  \bottomrule
\end{tabular}
\caption{The example of an advertisement int-abs pair: broad intents capture potential needs, precise abstract conveys product information.}
 \label{tab:example}
\end{table*}

\paragraph{Learning to Discriminate.} The goal of discrimination is to judge the relevance between the user's query and the int-abs generated by LLM to improve the accuracy of the model generation results. 
Specifically, we append a special token [CLS] at the end of the input prompt. Let the output representation of the model corresponding to this token be denoted as $h_{[CLS]} \in R^{d} $. We then apply a multi-layer perceptron (MLP) function f: $R^{d} \rightarrow R $ to $h_{[CLS]}$ to obtain a scalar relevance score:

\begin{equation}
%\begin{split}
s = f(h_{[CLS]}).
%\end{split}
\end{equation}

The ground-truth relevance labels, denoted by $y\in \{0,1\}$, are obtained from real online click-exposure logs. During training, we interpret the scalar score s as the logit for the predicted relevance probability via the sigmoid function:

\begin{equation}
%\begin{split}
\hat{p}  =   \sigma (s) =   \frac{1}{1+ e^{-s} } ,
%\end{split}
\end{equation}

where $\hat{p}$ represents the predicted probability that the input corresponds to a relevant sample. The training objective is to minimize the binary cross-entropy loss between the predicted relevance probability $\hat{p}$ and the ground-truth label $y$ , which is expressed as:

\begin{equation}
%\begin{split}
L_{dis}  =  - [ylog(\hat{p})+(1- y)log(1- \hat{p})].
%\end{split}
\end{equation}

%This loss function quantifies the discrepancy between the predicted relevance and the actual relevance labels derived from user feedback, and it is used to optimize the model parameters. 

\paragraph{Learning to Rank.}  The goal of the ranking stage is to prioritize ads with higher commercial value. Specifically, we adopt direct preference optimization (DPO) \cite{rafailov2023direct} to learn the relative ranking of ads. The training data consists of tuples $(x,y_{w}, y_{l})$, where $x$ denotes the context information, and $y_{w}$ and $y_{l}$ represent the winning (higher-performing) and losing (lower-performing) ads, respectively. The model assigns a generation probability to each ad $y$ conditioned on context $x$ via the policy $\pi_{\theta}$, while $\pi_{ref}$ is a reference policy. To measure the commercial difference and guide the model to learn the correct ranking, we define a loss function based on the ratio of probabilities as follows:

\begin{equation}\label{formula.3}
\begin{split}
L_{dpo} ( \pi_{\theta};\pi_{ref})  = -[log\sigma(\beta log \frac{ \pi_{\theta}(y_{w}|x) }{\pi_{ref}(y_{w}|x)} \\
- \beta log \frac{ \pi_{\theta}(y_{l}|x) }{\pi_{ref}(y_{l}|x)})]
\end{split}
\end{equation}

where $\sigma(\bullet)$ is the sigmoid function that maps the difference of log-probability ratios into the range (0,1) , reflecting the relative relevance. $\beta > 0$ is a scaling hyperparameter controlling the sharpness of this comparison. By minimizing this loss, the model is encouraged to assign significantly higher relative probabilities to winning ads than to losing ads, thereby achieving a ranking that prioritizes advertisements with greater commercial value.
%最终，我们将这三个损失函数加权，作为最终的模型损失函数进行训练，其中a,r,v均为超参数。

Finally, the overall loss of GDR-LLM is the sum of multiple loss functions, as shown in Formula~\ref{formula.4},

\begin{equation}\label{formula.4}
%\begin{split}
L_{model}  = \alpha L_{gen}  + \beta L_{dis} + \gamma L_{dpo}
%\end{split}
\end{equation}
where  ${\alpha ,  \beta ,  \gamma}$ are hyperparameters.

\paragraph{Training Data.}  
\label{traindata}
Given a user query q, we first collect online logs to obtain two sets of advertisements: a set of related ads with clicks A = \{${a_{1}}$...${a_{n}}$\} and a set of irrelevant ads B = \{${b_{1}}$...${b_{n}}$\} associated with q. We then train the generation model using the relevant set A by optimizing the loss ${L_{gen}}$. Meanwhile, the discriminator loss ${L_{dis}}$ and the ranking loss  ${L_{dpo}}$ are trained using both A and B. 

To generate candidate ads for ranking, we first take the union of sets A and B and sort them in descending order of commercial value, producing the candidate list C. This ordering reflects the ranking as predicted by the model trained with the ${L_{dpo}}$.

%To obtain candidate advertisements for ranking, we sort the union of A and B in descending order according to their commercial value, resulting in the candidate list C. This sorting reflects the ranking predicted by the model trained with   For training loss ${L_{dis}}$, candidate ads ${c_{i}}$ $\in$ A are assigned a label of 1, while ${c_{i}}$ $\in$ B are assigned a label of 0.

For each candidate ad  ${c_{i}}$ $\in$ C, we further assign relevance labels and determine the generation loss mask based on its set membership. Specifically, if ${c_{i}}$ $\in$ A, we set the generation loss mask to 1 and assign a relevance label of 1. If ${c_{i}}$ $\in$ B, the generation loss mask is set to 0 and the relevance label is 0. These relevance labels correspond to the correlations predicted by the model trained using ${L_{dis}}$. The final generation loss is calculated as the element-wise product of the mask and ${L_{gen}}$.

%For each candidate ad ${c_{i}}$ $\in$ C, if ${c_{i}}$ $\in$ A, we set the generation loss mask to 1 and assign the relevance label as 1. Conversely, if ${c_{i}}$ $\in$ B  the generation loss mask is set to 0 and the relevance label is 0.  This relevance label reflects the correlations predicted by the model trained using ${L_{dis}}$. The generation loss is then computed as the element-wise product of the mask and ${L_{gen}}$, i.e., mask * ${L_{gen}}$. 

\subsection{Ad Indexing} We employ LLM to generate commercial intents (SText) and ad abstracts (STerm). An inverted index mapping SText and STerm to ads is then built for efficient retrieval.

\label{int-abs}

\paragraph{Hierarchical Identifiers: Intent x Abstract.} In this work, each advertisement is represented by combining two complementary textual sources \textemdash broad intent and precise abstract. The semantic intent is a compressed text (SText) that provides a high-level overview of the commercial intent of the ad.  In contrast, the text abstract (STerm) is n-grams extracted from the ad display title, preserving specific factual details and salient phrases. By integrating these two components (SText x STerm), the joint representation leverages both the broad conceptual understanding offered by the semantic intent and the precise, grounded information from the extracted abstract. %Therefore, the fused representation, by integrating complementary semantic and extracted information, offers a more comprehensive and precise characterization of documents. This enhanced representation facilitates improved full recall and serves as a robust basis for accurately estimating both the relevance between queries and advertisements, as well as the commercial value of the advertisements. 

% 需要详细的解释一些这个example
Table~\ref{tab:example} presents an example of an ad’s commercial intent and abstract.  For a game title of Endless Winter, the intent refers to its broad commercial category or gameplay type, such as "Role-playing Game" or "Survival Challenge Game." In contrast, the abstract is limited to keywords that appear in the advertisement and are directly related to the title, such as "Endless Winter". Thus, "intent" captures the general purpose or genre, while "abstract" reflects the specific terms highlighted in ad.

\begin{figure}[t]
  \centerline{\includegraphics[width=1\linewidth]{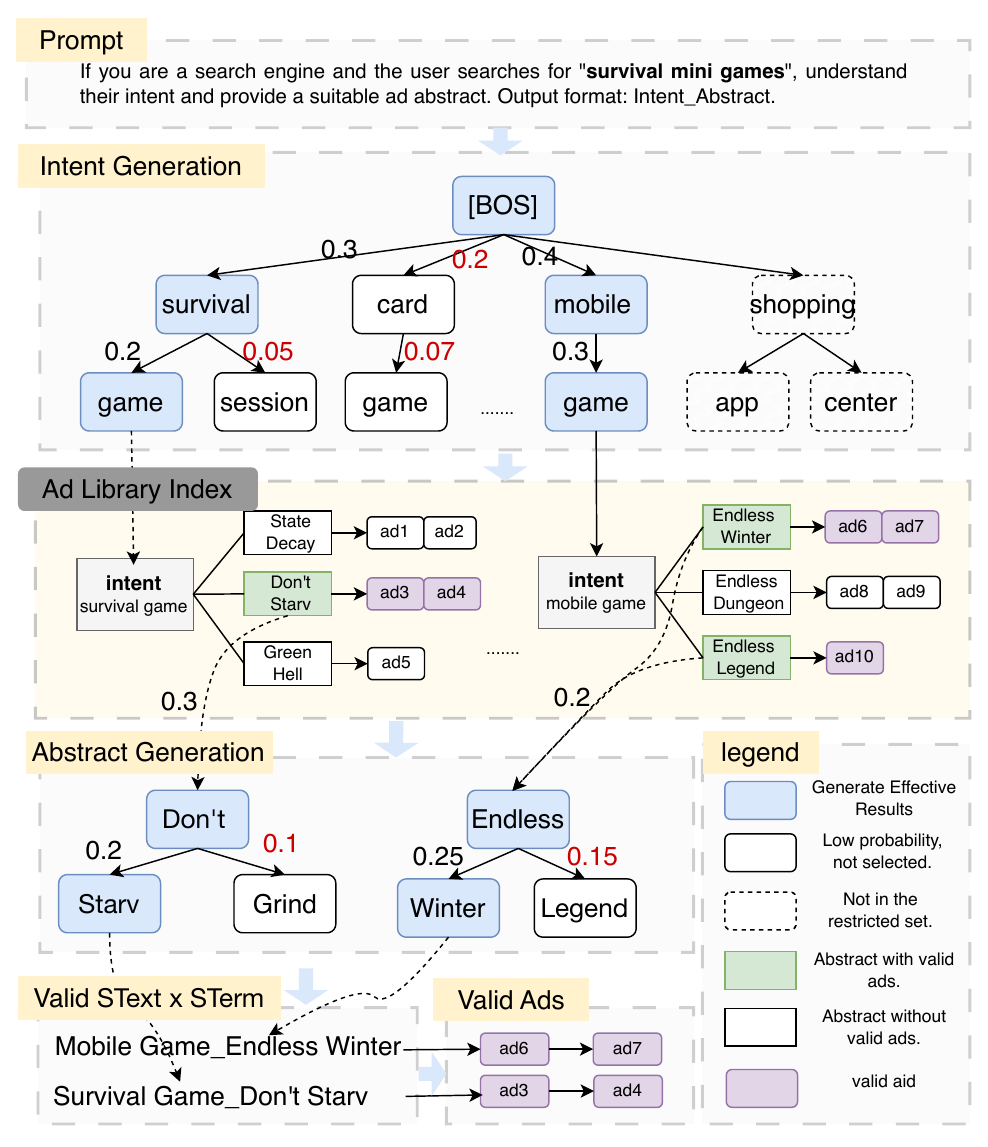}}
  \caption {Overview of dynamic constrained decoding: generate intent, filter invalid intent, and dynamical  reconstruct constrained trees for ad abstract generation.
  \label{fig:beam_search}
}
\end{figure}

% 这里写广告侧的意图是受限生成的，摘要是自由生成的
\paragraph{Ad Indexing} The ad indexing module aims to build an index that maps hierarchical representations to ads. We first use a large language model (LLM) to generate semantic intents for all ads in the database. These intents are deduplicated and cleaned at the industry level, producing approximately 700,000 coarse-grained commercial intent categories. A constrained beam search is then performed across the ad corpus over these intent categories, as shown in ad index construction module of Figure~\ref{fig:LOGR}. Next, the LLM generates a free (unconstrained) text abstracts for each ad, forming hierarchical representations of intent × abstract (int-abs) pair, which are used to construct an inverted index. For new ads, intents are generated under constraints, followed by unconstrained abstract generation. This enables efficient index updates without requiring the model to retrain or relearn the mapping from ad representations to ads.
%This hierarchical approach first classifies ads into broad intent categories, then refines representation through precise abstracts, supporting scalable categorization and fine-grained ad representation.

\begin{table*}[]
\centering
\begin{tabular}{c|c|lllll}
\toprule
\multicolumn{2}{c|}{\textbf{Method}}                                                           & \multicolumn{1}{c}{\textbf{HR@10}} & \multicolumn{1}{c}{\textbf{HR@50}} & \multicolumn{1}{c}{\textbf{HR@100}} & \multicolumn{1}{c}{\textbf{MAP}} & \multicolumn{1}{c}{\textbf{ACR}} \\ \midrule

%{Word-based}   & BM25   & 0.1342 & 0.2576 & 0.3382 & 0.4909 & 40.97\% \\ 
%\midrule
\multirow{2}{*}{Semantic-based}   & Bert          & 0.1313 & 0.2384 & 0.3075 & 0.3910 & 47.70\% \\
                                  & SimBert-v2-R  & 0.1339 & 0.2265 & 0.2830 & 0.3966 & 47.71\% \\ 
\midrule
\multirow{2}{*}{\begin{tabular}[c]{@{}c@{}} Generative \\ Retrieval\end{tabular}}  
                                  & SimBert-v2-G   &0.1294 & 0.2374 & 0.3095 & 0.3852 & 36.89\%    \\
                                  & T5-base             &0.0981 & 0.1880 & 0.2495 & 0.2910 & 29.15\%    \\ 
\midrule
\multirow{3}{*}{\begin{tabular}[c]{@{}c@{}} LLM-based \\ Generative \\ Retrieval \end{tabular}} 
                                  & Qwen-1.8B   & 0.0960  & 0.1887 & 0.2524 & 0.3138 & 47.27\%\\
                                  & Hunyuan-2B   & 0.1153 & 0.2184 & 0.2889 & 0.3519 &  42.51\% \\
                                  & DSI-1B &  0.0320 & 0.0830 & 0.1058  & 0.2831 & 55.49\% \\
\midrule
Ours   & \textbf{ANGLE-1B}        & \multicolumn{1}{l}{\textbf{0.1961}} & \multicolumn{1}{l}{\textbf{0.4526}} & \multicolumn{1}{l}{\textbf{0.5273}}  & \multicolumn{1}{l}{\textbf{0.4834}}  &  \multicolumn{1}{l}{\textbf{69.17\%}}   \\ 
\bottomrule

\end{tabular}
\caption{Comparison of ANGLE and Baseline Models in Offline Scenarios.}
\label{tab:eval-res}
\end{table*}

\begin{table*}
  \centering
  \begin{tabular}{c|ccccc}
    \toprule
   {\textbf{\begin{tabular}[c]{@{}c@{}}Online Scenarios\end{tabular}}} & \textbf{Consumption} & \textbf{GMV} & \textbf{Clicks} & \textbf{\begin{tabular}[c]{@{}c@{}}Conversions\end{tabular}} & \textbf{\begin{tabular}[c]{@{}c@{}}Exposure\end{tabular}} \\ 
    \midrule
    \verb|WTS(anonymous)|  & {+1.81\%}  & {+2.16\%}  & {+1.50\%} &{+1.44\%} &{+2.49\%}       \\
    %\verb|Demand-Side Platform|  & {+7.18\%} &{+5.03\%} &{-} &{+6.85\%} &{+5.93\%}   \\
     \verb|QBS(anonymous)|  & {+6.57\%} &{+5.01\%} &{+6.11\%} &{+0.72\%} &{+7.82\%}         \\
     \verb|IPS(anonymous)|  & {+3.02\%} &{+2.39\%} &{+2.55\%} &{+2.16\%} &{+9.84\%}         \\
     \bottomrule
  \end{tabular}
  \caption{Application of ANGLE to Real-world Search Systems.}
  \label{tab:ab-testing}
\end{table*}

\subsection{Retrieval}

\paragraph{Dynamical Constrained Beam Search (DCBS).} %介绍受限解码的具体流程
Dynamic constrained beam search leverages a dynamic tree-building module to ensure that all generated advertisements (ads) belong to a predefined valid set. As illustrated in Figure \ref{fig:beam_search} with a beam size of 2, the process begins by constructing a prefix trie of intent-abstract (int-abs) pairs. Generation starts from the root token ([BOS]) by producing intent tokens. During decoding, the ad library index module continuously verifies the validity of ads associated with each generated abstract linked to the intents. Only abstracts corresponding to valid ads are retained as candidates for further expansion, while those without valid ads are immediately pruned from the search space. If all abstracts under a given intent are invalid, that intent is also discarded, terminating its generation path. For the same query, different users have different sets of valid ads, resulting in different sets of valid int-abs pairs. Therefore, valid int-abs pairs are dynamically generated based on user features. This dynamic filtering effectively restricts decoding to valid int-abs pairs, ensuring that the generation process produces only valid ads in real time.

Typically, organizing the complete set of ads into a constrained tree ensures that every ad can potentially be retrieved during inference. However, many of these retrieved ads may be irrelevant or invalid for the specific user context. For instance, an ad targeted exclusively to female users would be ineffective if the current searcher is male. In practice, the retrieval system relies on target filtering within the MCA module (Figure \ref{fig:figure_2}) to exclude ads that do not meet delivery requirements. As a result, a large portion of generated ad candidates, despite satisfying relevance and commercial criteria, may ultimately prove ineffective. By contrast, DCBS directly integrates validity constraints into the generation process, dynamically pruning invalid int-abs pairs in real time. This approach not only guarantees that all generated candidates are valid for the target user but also substantially improves generation efficiency and practical effectiveness by minimizing wasted retrieval and filtering efforts downstream.

\paragraph{Retrieval.} This section mainly introduces the retrieval part of the ANGLE. As shown in the retrieval of Figure~\ref{fig:beam_search}, upon receiving a user query, the GDR-LLM first performs a constrained generation to produce a beam of candidate user search intents. The dynamical constrained beam search then filters out ad abstracts that do not correspond to any valid ads, retaining only those linked to available ads. These valid ad abstracts are reorganized into a constrained tree structure, on which the LLM continues generation. This ensures that every hierarchical output aligns with ads that are valid and ready for display. By dynamically focusing decoding on this refined set of candidates, the framework significantly improves efficiency. After generating all valid int-abs pairs, the discriminator head produces relevance scores between the query and each int-abs pair, which are used to assess the quality of the generated outputs. %Additionally, newly added ads can be easily incorporated by appending their abstracts under the appropriate intent node, allowing seamless and scalable index updates.

% 需要再进一步修改 todo
\section{Experiments}

In this section, we first present the experimental settings, including the training dataset, evaluation dataset, baselines, and implementation details. We then report the experimental results of ANGLE in both online and offline scenarios, followed by an ablation study to evaluate the contribution and effectiveness of each component within ANGLE.
\subsection{Experimental Settings}
\label{setting}
\begin{table*}
  \centering
  \begin{tabular}{c|ccccc}
    \toprule
   {\textbf{\begin{tabular}[c]{@{}c@{}}Method\end{tabular}}} & \textbf{HR@10} & \textbf{HR@50} & \textbf{HR@100} & \textbf{\begin{tabular}[c]{@{}c@{}}MAP\end{tabular}} & \textbf{\begin{tabular}[c]{@{}c@{}}ACR\end{tabular}} \\ 
    \midrule
    \verb|w/o. Intent|  & {0.1867}  & {0.4291}  & {0.4645} &{0.4618} &{68.77\%}   \\
    \verb|w/o. Abstract|  & {0.1738} &{0.3384} &{0.3519} &{0.4361} &{68.88\%}   \\
     \verb|w/o. DCBS|  & {0.1442} &  {0.2803} &{0.2975}  &{0.3130} &{55.14\%}  \\
     \verb|w/o. GDR-LLM|  & {0.1278} &{0.4329} &{0.4412} &{0.4275} &{62.62\%}  \\
     \textbf{ANGLE}  & {\textbf{0.1961}} &{\textbf{0.4526}} &{\textbf{0.5273}} &{\textbf{0.4834}} &{\textbf{69.17\%}}  \\
     \bottomrule
  \end{tabular}
  \caption{Ablation Study of ANGLE.}
  \label{tab:ablation-study}
\end{table*}

% 最好在附录给个具体的训练数据详情，这里描述的太简单了，例如初始训练数据来源于开源LLM模型，训练数据的组织形式
\paragraph{Training Dataset.} In this work, The LLMs used for queries and advertising, as well as their corresponding supervised fine-tuning (SFT) data, differ significantly. The ad-side model employs a base LLM, while the query-side model utilizes the GDR-LLM.  The advertising-side data primarily originates from a proprietary LLM (e.g., GPT-4, \citealp{achiam2023gpt}). Online advertising information, such as ad title and landing pages, is input into the large language model, which generates the int-abs pair of each ad. After cleaning, approximately 5,000 training samples were obtained. On the query side, the GDR-LLM is employed to regulate the relevance and commercial value of the generated outputs. Specifically, the SFT data is derived from online click logs. We collected click data from the past month and constructed a training dataset comprising 430,000 samples by incorporating negative examples. Subsequently, advertisements associated with each query were commercially ranked based on actual online expenditure to train the DPO model. Specific examples of the training data are presented in Table ~\ref{tab:appendix-1} of Appendix ~\ref{appendixA}, and the organization rules for the training data are described in the “Training Data” in section ~\ref{traindata}.

\paragraph{Evaluation dataset.} We collected online impression and click data spanning one month and selected 10k queries along with their corresponding clicked ads to form the evaluation dataset. For each query, up to 1k candidate ads were included. To increase the evaluation’s realism and challenge, a set of irrelevant ads was randomly sampled from the ad library and added to the candidate pool. As a result, the final evaluation dataset comprised approximately 220k ad candidates, ensuring both completeness and diversity for a thorough assessment.

\paragraph{Baselines.}  We compare ANGLE with seven benchmark models to validate its effectiveness.

Semantic-based methods use BERT \cite{devlin2019bert} or SimBert-v2 \cite{RoFormer-Sim} to generate semantic embeddings for queries and keywords, respectively, and apply approximate nearest neighbor (ANN) to retrieve keywords relevant to the query. Both BERT and SimBert-v2 models are based on a 12-layer transformer architecture with a hidden size of 768 and 12 attention heads, comprising approximately 110 million parameters. During training, we used online click logs as positive samples, and randomly sampled within the batch as negative examples.

Generative retrieval (T5, \citealp{raffel2020exploring}) and LLM-based generative retrieval (Qwen \cite{bai2023qwen}, HunYuan \cite{Tencent-Hunyuan}, DSI \cite{tay2022transformer} ) directly produce business keywords from the query. However, unconstrained decoding may generate keywords that do not correspond to any real ads. To address this issue, we adopt the hierarchical navigable small world (HNSW) \cite{malkov2018} to identify the most similar keyword within the existing database, which is then used as the final output instead of the original LLM-generated keywords.

\paragraph{Implementation Details.}

We use the Hunyuan-13B to generate hierarchical  text for ads. Because the online system has strict time constraints, the query uses the Hunyuan-1B, with a beam size set to 64. To ensure the stability of online system performance, the maximum length of the intent is limited to 3 tokens, while the abstract is restricted to 4 tokens. Hunyuan-1B requires approximately 8 milliseconds to generate each token; therefore, the total time for generating an int-abs pair is limited to within 90 milliseconds, meeting the latency requirements for online system. The ad index is updated hourly, allowing newly added ads to quickly generate int-abs pairs and update the index accordingly. Meanwhile, expired ads are promptly removed from the index.

\subsection{Experimental Results}

\paragraph{Online Result.} We conducted an A/B testing experiment over five days using 20\% of the traffic, which resulted in a significant increase of 1.81\% in consumption and 2.16\% in GMV. At the same time, both the number of conversions and clicks showed varying degrees of improvement, as presented in Table \ref{tab:ab-testing}. These results demonstrate that ANGLE effectively alleviates the issue of inconsistent target optimization in MCA. Currently, it has been fully applied to WTS (anonymous), QBS (anonymous) and IPS(anonymous) as a supplement to the ranking queue.

\paragraph{Offline Evaluation.}  

To assess the effectiveness of ANGLE, we constructed an evaluation dataset consisting of 10,000 queries and compared its performance with seven strong baselines. The results, summarized in Table ~\ref{tab:eval-res}, demonstrate that ANGLE achieves state-of-the-art (SOTA) performance across all evaluation metrics, including HR, MAP, and ACR.

% todo
\paragraph{Ablation Study.} 
We validated the effectiveness of ANGLE through ablation experiments, as shown in Table~\ref{tab:ablation-study}, where "w/o. Intent" refers to representing ads solely using abstracts; "w/o. Abstract" refers to representing ads solely using commercial intents. "w/o. DCBS" refers to removing the DCBS —allowing the LLM to freely generate the int-abs pairs for ads and subsequently using ANN to obtain similar int-abs pairs for the purpose of retrieving ads. "w/o. GDC-LLM" refers to utilize a base LLM to perform constrained generation. The ablation results demonstrate that removing any module negatively impacts the overall performance.

\section{Conclusion}

%In this work, we present ANGLE, a unified end-to-end generative framework for real-time ad retrieval using a single large language model. ANGLE integrates ad generation, discrimination, and ranking, effectively resolving issues of inconsistent objective optimization and error propagation in traditional multi-stage systems. By leveraging hierarchical text representations, ANGLE reduces reliance on memorizing SID-ad pairs and improves generalization, adaptability, and robustness. Extensive online A/B tests show that ANGLE increases consumption by 1.81% and GMV by 2.16%, demonstrating its practical effectiveness and potential for LLM-driven advertising solutions.

This work introduces ANGLE, a unified end-to-end generative framework for real-time ad retrieval. ANGLE integrates ad generation, discrimination, and ranking into a single LLM, enabling direct access to final ranking for ads. This effectively addresses the common issues of logical inconsistency and error propagation in MCA. Hierarchical text representations further alleviate generalization challenges related to SID and reduce reliance on memorizing SID-Ad pairs, improving adaptability and robustness. Extensive evaluations, including online A/B tests, offline experiments, validate ANGLE’s effectiveness and highlight its strong potential for advancing LLM-based advertising solutions.

\section*{Limitations}

This section discusses the limitations of the current work. Unlike recommendation systems, retrieval systems involve complex components such as dynamic creative selection and intelligent bidding, all of which rely heavily on ranking. Consequently, our proposed model does not eliminate the use of ranking. Future research will investigate the feasibility of removing precise ranking from search systems, aiming to further leverage the potential of large-scale models.

\bibliography{custom}

\newpage
\appendix

\section{Problem Definition}
\label{appendixA}
In search advertising, given a user query q  and a large candidate ad set A = \{$a_{1},a_{1}...a_{n}$\}, where n can reach tens of millions, the goal of the retrieval system is to select ads that satisfy user intent and maximize the platform’s commercial value. Traditionally, this involves a five-stage cascaded pipeline: retrieval, relevance, pre-ranking, and ranking, where each stage is optimized independently. This leads to inconsistent objectives and error propagation, degrading overall performance. To overcome these issues, we propose a unified end-to-end framework that integrates retrieval, relevance, and ranking to directly generate the final set of high-quality ads, thereby minimizing inter-stage discrepancies and cumulative errors.  For GDR-LLM, training data are constructed using online logs. Specifically, for each query, we collect advertisements that have been clicked by users in the online environment and retrieve their corresponding intent-abstract (int-abs) pairs. By associating each query with the int-abs pairs of the ads it led users to click, we create a high-quality training dataset consisting of query and int-abs pairs. This data construction method leverages real user interactions to provide relevant and diverse supervision signals for model fine-tuning, thereby improving the model’s ability to understand query-advertisement relationships in practical scenarios.

\section{Fine-tuning Data}

This section primarily describes the datasets used for fine-tuning the GDR-LLM on the query side and the base LLM on the advertisement side. As shown in Table 5, the training data for the base LLM are derived from a proprietary LLM such as GPT-4. For each advertisement title, the model is first prompted to generate the advertising intent, followed by a request to produce a concise abstract of the advertisement. The specific prompts used in this process are detailed in Table 5. Subsequently, both the generated intent and abstract undergo rigorous data cleaning, and are then combined to create multiple intent-abstract pairs corresponding to each advertisement title. This structured approach ensures that the training data are high-quality and relevant, thereby enhancing the model’s performance in downstream tasks.

\section{Evaluation Metrics}
In this paper, ACR~\cite{fan2019mobius}, Hit Ratio (HR@K)~\cite{alsini2020hit}, and Mean Average Precision (MAP) ~\cite{cormack2006statistical} are used to evaluate the performance of ANGLE.
\textit{Ad Coverage Rate} (ACR) is a metric used to evaluate the proportion of unique advertisements that a system is able to retrieve  from the total pool of available ads. A higher ACR indicates that the retrieval system can expose users to a broader variety of ads, which is important for both user experience and advertiser fairness. Ad Pave View (AdPV) is the number of requests with ad recall, and Pave View (PV) is the number of requests.

\begin{equation}\label{formula.20}
{ACR = AdPV/PV}
\end{equation}

\textit{Hit Ratio (HR@K)} is a metric that measures the proportion of relevant items appearing in the top K results of a  retrieval system. Specifically, Hits@K represents the number of relevant ads within the top-K retrieved candidates that belong to the ground truth set and Ground Truth (GT) represents set of candidate ads. A higher HR@K reflects better retrieval accuracy and effectiveness for users.

\begin{equation}\label{formula.21}
{HR@K =  \frac{Hits@K}{|GT|} }
\end{equation}

\textit{Mean Average Precision (MAP)} measures overall ranking quality by averaging the Average Precision (AP) across all queries, as defined in Equation~\ref{formula.23}.

\begin{equation}\label{formula.23}
MAP =  \frac{1}{Q} \sum_{q \in Q} AP_{q}
\end{equation}

\textit{Average Precision (AP)} for a given query is described in Equation~\ref{formula.24}. In this formula, $\Omega_{q}$ denotes the set of true relevant items, $p_{qj}$ is the rank position of $ad_{j}$ in the output list, and $p_{qj} < p_{qi}$ signifies that $ad_{j}$ is ranked higher than $ad_{i}$.

\begin{equation}\label{formula.24}
AP_{q} = \frac{1}{|\Omega_{q}|} \sum_{i \in \Omega_{q}} \frac{\sum_{j \in \Omega_{q}} h(p_{qj} < p_{qi}) + 1}{p_{qi}}
\end{equation}

\section{Online Inference Support}
For online inference deployment, we have constructed a dedicated GPU cluster comprising hundreds of L40 GPUs. This cluster has been optimized for workload distribution and peak GPU utilization, achieving a utilization rate of nearly 90\%. The trained models are converted to FP8 format, enabling each L40 GPU to process approximately 30 queries per second. We also optimize each stage of the inference pipeline, including input preprocessing, model inference, output postprocessing, and network transmission, in order to reduce overall end-to-end latency and ensure that each request can be completed within 90 ms.

\begin{table*}
  \centering
  \begin{tabular}{m{2cm}<{\centering}m{6cm}m{6cm}m{1cm}<{\centering}}
    \hline
    \textbf{Model Used} &  \textbf{Prompt} &  \textbf{Output} & \textbf{Data Size} \\
    \hline
    \begin{tabular}[c]{@{}c@{}}\textbf{GDR-LLM} \\ query-side\end{tabular}   & \begin{CJK}{UTF8}{gbsn} \small{As a search system, when a user enters a query = "12306", please first analyze the user's search intent. Then, recommend suitable advertisements based on that intent, and generate corresponding ad abstract information for display.} \end{CJK} & \begin{CJK}{UTF8}{gbsn}\small{\textbf{From online click logs:}Train Ticket Booking\_12306 Official; High-Speed Rail Ticket\_12306 Official; Online Ticketing\_12306 Official; Travel Services\_12306 Official....... }\end{CJK} & 430k \\ \hline
    \multirow{2}{*}{\begin{tabular}[c]{@{}c@{}} \textbf{Base LLM} \\ads-side \end{tabular}}  & \begin{CJK}{UTF8}{gbsn} \small{
    \textbf{Task Requirements:} Generate advertising titles based on user search queries and extract a abstract for each title. The abstract should follow the format: Commercial Intent\_Brand/Core Service, where "Commercial Intent" specifies a concrete product or service category, and "Brand/Core Service" clearly identifies the brand or core service. Correct Example: Organic Vegetables\_Meituan Grocery (product category + brand), Incorrect Example: Organic Vegetables\_Direct from Source ("Direct from Source" is a selling point, not a brand or service).   \textbf{ Business Point Guidelines}:  Must directly reflect the user's search intent (e.g., “Fresh Food Delivery,” “Grocery Shopping App”) Avoid vague terms (e.g., “Quality Guaranteed,” “Fast Delivery”).  \textbf{abstract Guidelines}: Must include a brand name (e.g., "Meituan Grocery") or core service name (e.g., "Gaming").  Given the query “12306”, please perform the following steps: 1. Identify specific business points from the query, ensuring alignment with the above criteria. 2.Generate ad titles conveying commercial information for each business point. 3.Extract a compliant abstract from each title. 4.Combine each business point with its abstract in the format: Business Point\_Brand/Core Service. 5.On the last line, concatenate all results separated by semicolons.} \end{CJK}  & \begin{CJK}{UTF8}{gbsn}  \small{\textbf{Output from LLM models (e.g. GPT-4): } Step-by-step execution: 
    1. Business point analysis:
    * Train ticket booking (User need: purchasing train tickets/high-speed rail tickets)
    * High-speed rail ticket discounts (User need: purchasing tickets at lower prices)
    * Online ticketing platforms (User need: booking tickets through official channels)
    * Travel services (User need: train schedule inquiries/ticket availability monitoring)
    2. Advertisement headline generation:
    * Train ticket booking: Official 12306 booking: High-speed rail tickets starting from 50\% off, online seat selection with no service fees!
    * High-speed rail ticket discounts: 12306 limited-time offer! All student/adult tickets at 20\% off
    * Online ticketing platforms: 12306 official booking: Real-time ticket availability + standby ticket grabbing, click to purchase
    * Travel services: 12306 official services: Train schedules, ticket availability monitoring, worry-free travel!
    3. Title summarization and extraction:
    * Train ticket booking : Train Ticket Booking\_12306 Official
    * High-speed rail ticket discounts : High-Speed Rail Ticket\_12306 Official
    * Online ticketing platform : Online Ticketing\_12306 Official
    * Travel services : Travel Services\_12306 Official
    Final output:
    Business Point\_Brand/Core Service:
    Train Ticket Booking\_12306 Official; High-Speed Rail Ticket\_12306 Official; Online Ticketing\_12306 Official; Travel Services\_12306 Official
    } \end{CJK}& 5k \\  \hline

  \end{tabular}
  \caption{Details of fine-tuning data.}
  \label{tab:appendix-1}
\end{table*}

%  

% \usepackage{multirow}
%\section{Example Appendix}
%\label{sec:appendix}

%This is an appendix.

\end{document}